\newcommand{\etal}{{\it et al.}}
\documentclass[twocolumn,twoside,slac_two]{revtex4}
\usepackage{graphicx}
\usepackage{fancyhdr}
\renewcommand{\headrulewidth}{0pt}
\renewcommand{\footrulewidth}{0pt}

\begin{document}

\title{The Purely Leptonic Decays \boldmath{$D^+\to \mu^+\nu$} and \boldmath{$D_s^+\to \ell^+\nu$} at CLEO}

%

\author{Sheldon Stone}
\affiliation{Physics Department, Syracuse University, Syracuse, N. Y. 13244, USA}

\begin{abstract}
We update our previous results by increasing
the luminosity, the efficiency, and for the $D_s^+$ the number of
tags. We determine $f_{D^+}=(205.8\pm 8.5
\pm 2.5)$ MeV, and an interim preliminary value
 of $f_{D_s^+}=(267.9\pm 8.2\pm 3.9)$ MeV, where both results are
 radiatively corrected. We agree with the recent
 most precise unquenched Lattice-QCD calculation for the $D^+$, but
 are in disagreement for the $D_s^+$. Several
 consequences are discussed, including the possibility of physics beyond
 the Standard Model.
\end{abstract}

\maketitle

\thispagestyle{fancy}


\section{Introduction}
\begin{figure}[h]
\centering
\includegraphics[width=80mm]{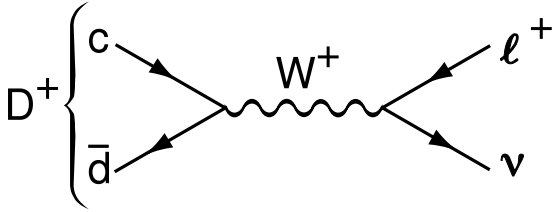}
\caption{The
decay diagram for $D^+\to \ell^+\nu$.} \label{Dptomunu}
\end{figure}

Purely leptonic decays of heavy mesons proceed in the Standard Model
(SM) via a $W^+$ annihilation diagram shown specifically for $D^+\to
\ell^+\nu$ in Fig.~\ref{Dptomunu}.  The strong interaction effects
are parameterized in terms of the ``decay constant" for the $D^+$
meson $f_{D^+}$.  The decay width is given by
\begin{equation}
\Gamma(D^+\to \ell^+\nu) = {{G_F^2}\over
8\pi}f_{D^+}^2m_{\ell}^2M_{D^+} \left(1-{m_{\ell}^2\over
M_{D^+}^2}\right)^2 \left|V_{cd}\right|^2~~, \label{eq:equ_rate}
\end{equation}
where $G_F$ is the Fermi constant, $M_{D^+}$ is the $D^+$ mass,
$m_{\ell}$ the final state charged-lepton mass, and $V_{cd}$ is a
CKM matrix element, taken equal to $V_{us}$. Thus, in the SM
measurement of purely leptonic decays allow a determination the
decay constant, $f_{D^+}$ of the $D^+$ meson and similarly $f_{D_s}$
of the $D_s^+$ meson.

Meson decay constants in the $B$ system are used to translate
measurements of $B\bar{B}$ mixing to CKM matrix elements. Currently,
it is not possible to determine $f_B$ accurately from leptonic $B$
decays, so theoretical calculations of $f_B$ must be used. Since the
$B^0_s$ meson does not have $\ell^+\nu$ decays, it will never be
possible to determine $f_{B_s}$ experimentally, so again theory must
be relied upon. If calculations disagree on $D$ mesons, they may be
questionable on $B$ mesons. If, on the other hand new physics is
present, it is imperative to understand how it effects SM based
predictions of the $B$ decay constants. Decay constants can be
calculated using Lattice-QCD techniques. Recently, Follana~\etal
~using an unquenched lattice technique predicted $f_{D^+}=(207\pm
4)$ MeV and $f_{D_s}=(241\pm 3)$ MeV. \cite{Lat:Follana}

In these analyses we exploit the reactions $e^+e^-\to D^-D^+$, and
$e^+e^-\to D_s^{*-}D_s^+$ or $D_s^{-}D_s^{*+}$. The $D^+$ is studied
at 3770 MeV using 818 pb$^{-1}$. $D_s^+$ is studied at 4170 MeV,
using 400 pb$^{-1}$  for the $\mu^+\nu$, and $\tau^+\nu$,
$\tau^+\to\pi^+\overline{\nu}$ final states, and 300 pb$^{-1}$ for
$\tau^+\to e^+\nu\overline{\nu}$. (Eventually CLEO will present results
using 600 pb$^{-1}$.)

\section{\boldmath $D^+\to \ell^+\nu$}

\begin{figure}[h]
\centering
\includegraphics[width=80mm]{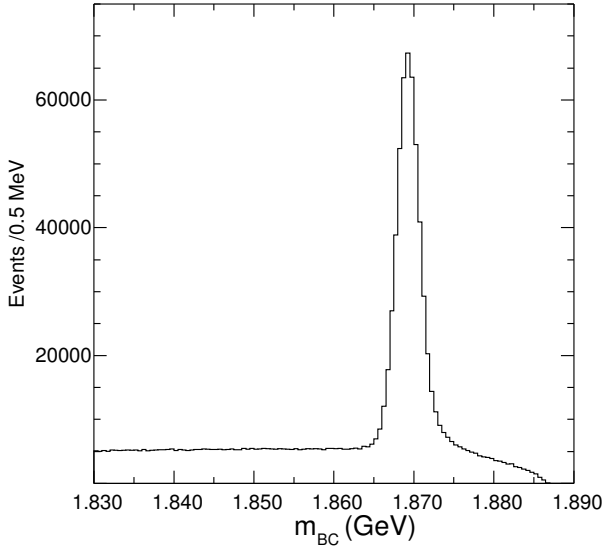}
\caption{The beam-constrained mass distributions summed over
$D^-$ decay candidates in the final states: $K^+ \pi^- \pi^-$, $K^+
\pi^- \pi^- \pi^0$, $K_S\pi^-$, $K_S \pi^-\pi^-\pi^+$, $K_S\pi^-
\pi^0$ and $K^+K^-\pi^-$.} \label{mbc_all}
\end{figure}

We use a ``double tag" technique where one $D^{\pm}$ is fully
reconstructed and the oppositely charged $D$ can then be found even
if there is a missing neutrino in the final state \cite{FinalDp}. For notational
convenance, the $D^-$ is referred to for the fully reconstructed
tag, although $D^+$ states are also used. To reconstruct $D^-$ tags
we require that the tag candidates have a measured energy consistent
with the beam energy, and have a ``beam constrained mass," $m_{BC}$,
consistent with the $D^-$ mass, where $m_{\rm BC}=\sqrt{E_{\rm
beam}^2-(\sum_i{\bf p}_{i})^2},$ $E_{\rm beam}$ is the beam energy
and $i$ runs over all the final state particles three-momenta.
Fig.~\ref{mbc_all} shows the $m_{BC}$ distribution summed over
all the decay modes we use for tagging. Selecting events in the mass
peak we count 460,055$\pm$787 signal events over a background of
89,472 events.

To search for signal events we look for events with one additional
track with opposite sign of charge to the tag. The track must have
an angle $>$25.8$^{\circ}$ with the beam line. We separate these
events into two categories. Case (i): those which deposit $<$ 300
MeV of energy in the calorimeter, characteristic of 99\% of muons,
and case (ii) those which deposit $>$ 300 MeV, characteristic of
45\% of the pions, those that happen to interact in the calorimeter.

We look for $D^+\to\mu^+\nu$ by computing the square of the missing
mass
\begin{equation}
{\rm MM}^2=\left(E_{\rm beam}-E_{\mu^+}\right)^2-\left(-{\bf
p}_{D^-} -{\bf p}_{\mu^+}\right)^2, \label{eq:MMsq}
\end{equation}
where ${\bf p}_{D^-}$ is the three-momentum of the fully
reconstructed $D^-$, and $E_{\mu^+}({\bf p}_{\mu^+})$ is the energy
(momentum) of the $\mu^+$ candidate. The signal peaks at zero for
$\mu^+\nu$ and is smeared toward more positive values for
$\tau^+\nu$, $\tau^+\to\pi^+\overline{\nu}$.

\begin{figure}[h]
\centering
\includegraphics[width=80mm]{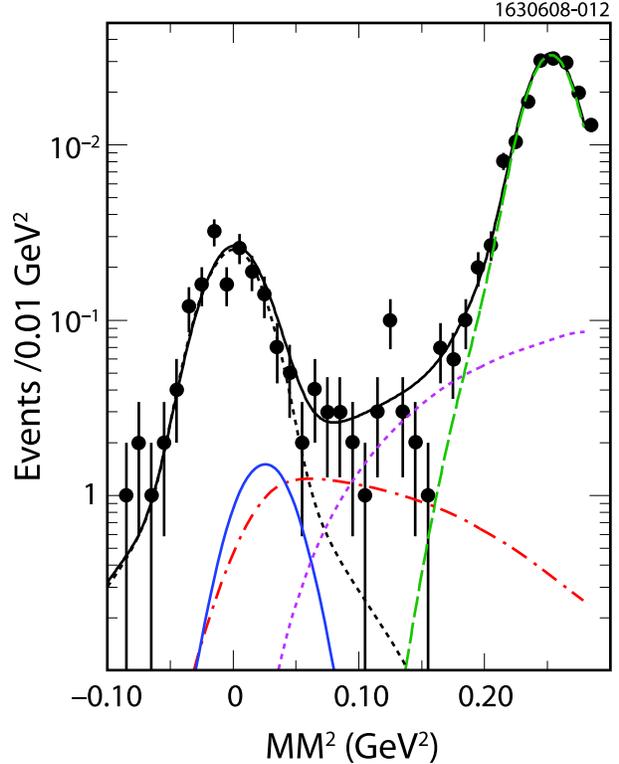}
\caption{Fit to the MM$^2$ for case (i). The
points with error bars
 show the data. The black (dashed) curve centered at zero
 shows the signal $\mu^+\nu$ events. The dot-dashed (red) curve that peaks around
 0.05 GeV$^2$ shows the $D^+\to\tau^+\nu$, $\tau^+\to\pi^+\bar{\nu}$
 component.  The solid (blue) Gaussian shaped
 curve centered on the pion-mass squared shows the residual
 $\pi^+\pi^0$ component. The dashed (purple) curve that falls to
 zero around 0.03 GeV$^2$ is the
 sum of all the other background components, except the $\overline{K}^0\pi^+$
 tail which is shown by the long-dashed (green) curve that peaks up at
 0.25 GeV$^2$. The solid (black) curve is the sum of all the
 components.} \label{case1-taunufix}
\end{figure}

 The fit to the
case (i) MM$^2$ distribution is shown in Fig.~\ref{case1-taunufix}
contains separate shapes for signal, $\pi^+\pi^0$,
$\overline{K}^0\pi^+$, $\tau^+\nu$ ($\tau^+\to \pi^+\bar{\nu})$, and
a background shape describing three-body decays. Here we assume the
SM ratio of 2.65 for the ratio of the $\tau^+\nu/\mu^+\nu$ component
and constrain the area ratio of these components to the product of
2.65 with ${\cal{B}}(\tau^+\to
\pi^+\bar{\nu}$)=(10.90$\pm$0.07)\%~\cite{PDG} and the 55\%
probability that the pion deposits $<$300 MeV in the calorimeter. We
veto events with an extra neutral energy cluster $>$ 250 MeV. This
removes most  $\pi^+\pi^0$ events; the residual background is fixed
at 9.2 events. The normalizations of the signal,
$\overline{K}^0\pi^+$, and 3-body background shapes are allowed to
float.

The fit yields 149.7$\pm$12.0 $\mu^+\nu$ signal events and 25.8
$\tau^+\nu$, $\tau^+\to\pi^+\bar{\nu}$ events (for the entire MM$^2$ range). We also perform
the fit allowing the $\tau^+\nu$, $\tau^+\to\pi^+\bar{\nu}$
component to float.  Then we find
153.9$\pm$13.5 $\mu^+\nu$ events and 13.5$\pm$15.3 $\tau^+\nu$,
$\tau^+\to\pi^+\bar{\nu}$ events, compared with the 25.8 we expect
in the SM. Performing the fit in this manner gives a
result that is independent of the SM expectation of the
$D^+\to\tau^+\nu$ rate. To extract a branching fraction, in either
case, we subtract off  2.4$\pm$1.0  events found from simulations and other studies to
be additional backgrounds, not taken into account by the fit.

We find ${\cal{B}}(D^+\to\mu^+\nu)=(3.82\pm 0.32\pm 0.09)\times
10^{-4}.$ The decay constant $f_{D^+}$ is then obtained from
Eq.~(\ref{eq:equ_rate}) using 1040$\pm$7 fs as the $D^+$ lifetime
\cite{PDG} and 0.2256 as $|V_{cd}|$. Our final result, including
radiative corrections is
\begin{equation}
f_{D^+}=(205.8\pm 8.5\pm 2.5)~{\rm MeV}~.
\end{equation}

\section{\boldmath $D_s^+\to \ell^+\nu$}

\begin{figure}[h]
\centering
\includegraphics[width=80mm]{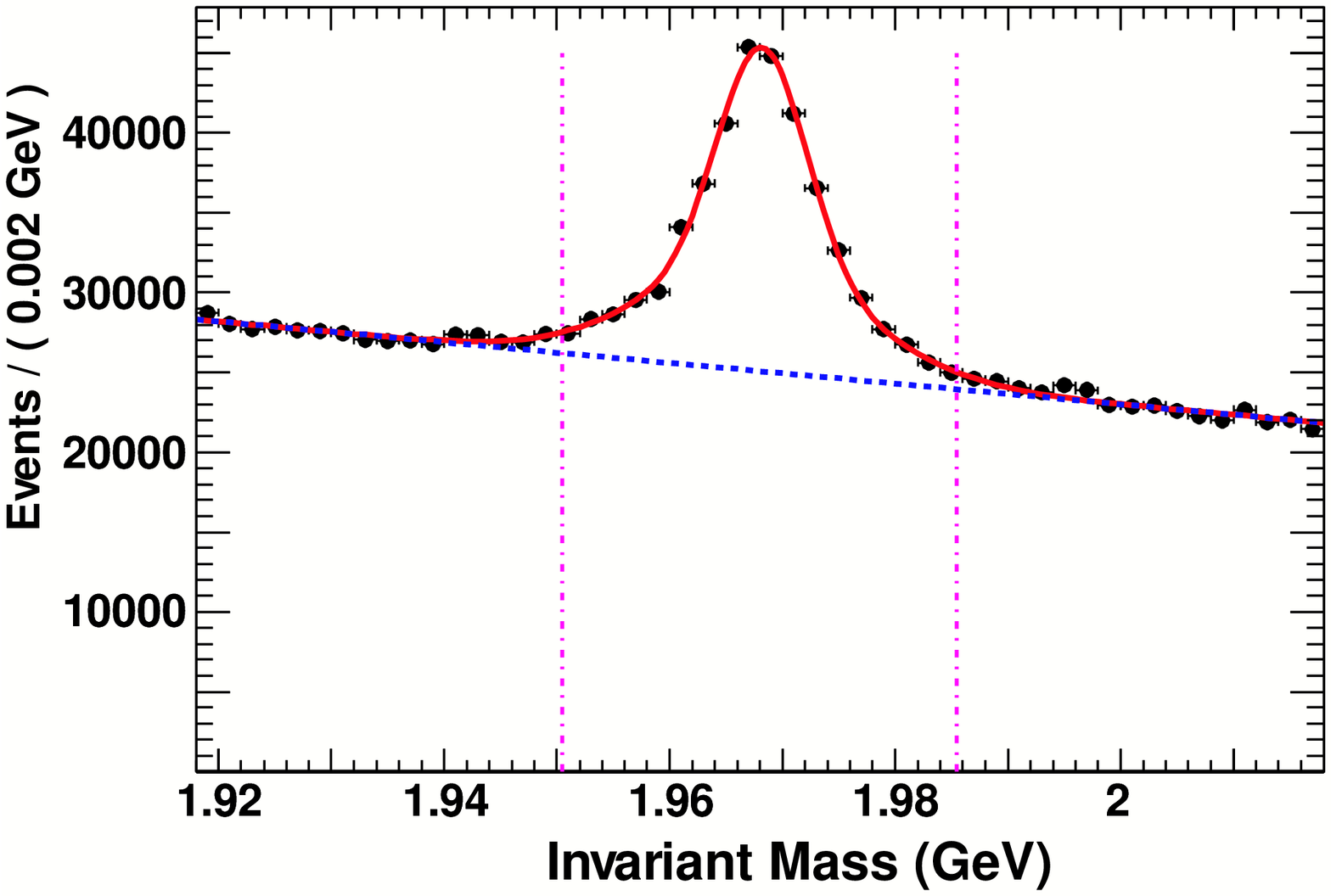}
\caption{The
invariant mass distributions summed over $D_s^-$ decay candidates in
the final states: $K^+K^-\pi^- $, $K_S K^-$, $\eta\pi^-$;
$\eta\to\gamma\gamma$, $\eta'\pi^-$; $\eta'\to\pi^+\pi^-\eta$,
$\eta\to\gamma\gamma$, $\phi\rho^-$; $\phi\to K^+K^-$, $\rho^-\to
\pi^-\pi^0$, $\pi^+\pi^-\pi^-$, $K^{*-}K^{*0}$; $K^{*-}\to
K_S^0\pi^-$, ${K}^{*0}\to K^+\pi^-$, $\eta\rho^-$;
$\eta\to\gamma\gamma$, $\rho^-\to \pi^-\pi^0$, and $\eta'\pi^-$;
$\eta'\to\pi^+\pi^-\gamma$. The curves represent signal and
background.} \label{mass38-47_all}
\end{figure}

In the $D_s$ case we have to take into account the additional photon
from $D_s^*\to\gamma D_s$, since we use $e^+e^-\to D_s^*D_s$ events.
We first examine the invariant masses (see
Fig.~\ref{mass38-47_all}). Then to select the appropriate $D_s^-$
tag and $\gamma$ candidate, we compute the square of the missing
mass opposite the selected tag and candidate $\gamma$'s, which peaks
at the $D_s^+$ mass for correct combinations, where
\begin{equation}
\label{eq:mmss} {\rm MM}^{*2}=\left(E_{\rm
CM}-E_{D_s}-E_{\gamma}\right)^2- \left({\bf p}_{\rm
CM}-{\bf p}_{D_s}-{\bf p}_{\gamma}\right)^2,
\end{equation}
here $E_{\rm CM}$ (${\bf p}_{\rm CM}$) is the center-of-mass energy
(momentum), $E_{D_s}$ (${\bf p}_{D_s}$) is the energy (momentum) of
the fully reconstructed $D_s^-$ tag, and $E_{\gamma}$ (${\bf
p}_{\gamma}$) is the energy (momentum) of the additional photon.  In
performing this calculation we use a kinematic fit that constrains
the decay products of the $D_s^-$ to the known $D_s$ mass and
conserves overall momentum and energy.

\begin{figure}[h]
\centering
\includegraphics[width=80mm]{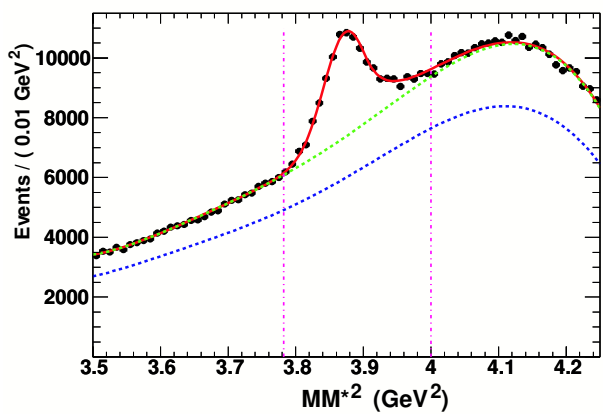}
\caption{The MM$^{*2}$ from a $\gamma$ plus tag candidate.
The curves represent signal and background components.} \label{mms238-47_sig}
\end{figure}

The MM$^{*2}$ distributions from the selected $D_s^-$ event sample
are shown in Fig.~\ref{mms238-47_sig}. We fit these distributions to
determine the number of tag events.

\begin{figure}[h]
\centering
\includegraphics[width=80mm]{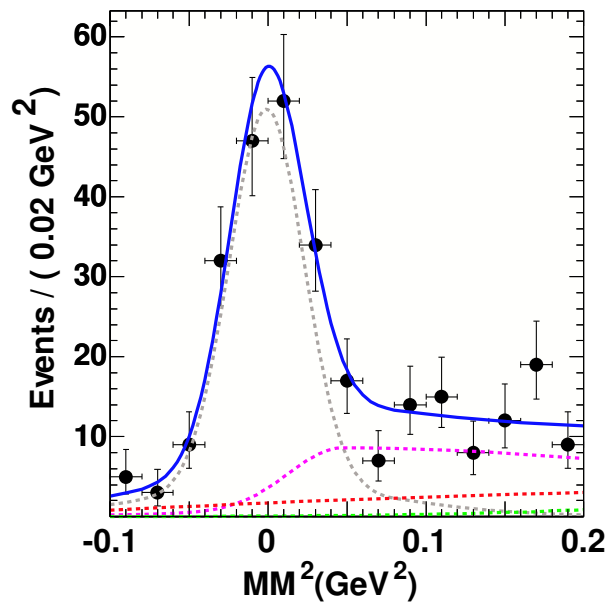}
\caption{The
MM$^2$ distribution. The dashed (grey) Gaussian shaped curve peaked
near zero is the $\mu^+\nu$ component, while the dashed (purple)
curve that rises sharply from zero and then flattens out shows the
$\tau^+\nu$ component. The two lines are background components. The
solid curve shows the sum. } \label{mms-fitall_38-47}
\end{figure}

After selecting the tags we then find events with a single
oppositely charged track to the tag and compute

\begin{eqnarray}
\label{eq:mm2} {\rm MM}^2&=&\left(E_{\rm
CM}-E_{D_s}-E_{\gamma}-E_{\mu}\right)^2\\\nonumber
           &&-\left({\bf p}_{\rm CM}-{\bf p}_{D_s}
           -{\bf p}_{\gamma}
           -{\bf p}_{\mu}\right)^2~.
\end{eqnarray}
We make use of a set of kinematical constraints and fit each event
to two hypotheses one of which is that the $D_s^-$ tag is the
daughter of a $D_s^{*-}$ and the other that the $D_s^{*+}$ decays
into $\gamma D_s^+$, with the $D_s^+$ subsequently decaying into
$\mu^+\nu$. The MM$^2$ distributions from data are shown in
Fig.~\ref{mms-fitall_38-47} where we have summed cases (i) and (ii).
After fixing the ratio of $\tau^+\nu/\mu^+\nu$ to the SM value we
find  $f_{D_s^+}=(268.2\pm 9.6 \pm4.4)$ MeV.

We can also use the decay mode $\tau^+\to e^+\nu\overline{\nu}$.
This result has already been published. \cite{CLEOtaunu} The
technique here is to use only three tagging modes: $\phi\pi^-$,
$K^-K^{*0}$ and $K^0_SK^-$, to ensure that the tags are extremely
clean. Then events with an identified $e^+$ and no other charged
tracks are selected. Any energy not associated with the tag decay
products is tabulated. Those events with small extra energy below
400 MeV are mostly pure $D_s^+\to\tau^+\nu$ events. After correcting
for efficiencies and residual backgrounds we find $f_{D_s^+}=(273\pm
16 \pm 8)$ MeV.

\section{Conclusions}

The preliminary CLEO average is $f_{D_s^+}=(267.9\pm 8.2 \pm 3.9)$
MeV (radiatively corrected). Averaging in the Belle result
\cite{Belle-munu} $f_{D_s^+}=(269.6\pm 8.3)$ MeV, which differs from
the Follana \etal~calculation~\cite{Lat:Follana} by 3.2 standard
deviations, while the result for $f_{D^+}=(205.8\pm 8.5 \pm 2.5)$
MeV is in good agreement. This discrepancy can be explained either
by New Physics \cite{Dobrescu-Kron} or casts suspicion on the
theoretical prediction. As similar calculations are used for
$f_{B_s}/f_B$, we need worry about them, or the effects of New
Physics on this ratio.

\section*{Acknowledgments}
I thank the U. S. National Science Foundation for support. Excellent
conversations were had with C. Davies, A. Kronfeld, P. Lepage,  P.
Mackenzie, J. Rosner R. Van de Water, and L. Zhang.


\begin{thebibliography}{99}

\bibitem{Lat:Follana}
E. Follana \etal ~(HPQCD and UKQCD), Phys. Rev. Lett.
{\bf 100}, 062002 (2008); see also C. Aubin \etal ~(FNAL Lattice, HPQCD \& MILC), {Phys. Rev. Lett.} {\bf 95}, 122002 (2005).

\bibitem{FinalDp}
For more details on this analysis see
B. I. Eisenstein \etal~(CLEO), 	arXiv:0806.2112 [hep-ex].

\bibitem{Rosner-Stone}
J. L. Rosner and S. Stone, ``Decay Constants of Charged Pseudoscalar
Mesons," to appear in PDG 2008, arXiv:0802.1043 [hep-ex].

\bibitem{Dobrescu-Kron}
B. A. Dobrescu and A. S. Kronfeld, arXiv:0803.0512 [hep-ph].


\bibitem{PDG}
W.-M. Yao \etal., {Journal of Physics,} {\bf G33}, 1 (2006).

\bibitem{CLEOtaunu}
K. M. Ecklund \etal~(CLEO), Phys. Rev. Lett. {\bf 100}, 161801
(2008).

\bibitem{Belle-munu} K. Abe \etal~(Belle),
 [arXiv:0709.1340] (2007).

\end{thebibliography}
\end{document}